\documentclass[reprint, amsmath,amssymb,aps,prd,superscriptaddress] {revtex4-2}

\usepackage{graphicx} 
\usepackage{dcolumn} 
\usepackage{bm} 
\usepackage{xspace}
\newcommand{\fuka}{\texttt{FUKA}\xspace}
\newcommand{\kadath}{\texttt{Kadath}\xspace}

\begin{document}
\title{Combining (post-)Newtonian ideas with quasi-equilibrium \\(QE) sequence analysis for black hole-neutron star (BHNS)\\ gravitational wave events.}
\author{Antonio Lanza}
\affiliation{SISSA, via Bonomea 265, 34136 Trieste, Italy}
\author{ Samuel D. Tootle}
\affiliation{Department of Physics, University of Idaho, Moscow, Idaho 83844, USA}
\author{Andrea Lapi}
\affiliation{SISSA, via Bonomea 265, 34136 Trieste, Italy}
\date {\today}

	\begin{abstract}
	In this paper we present quasi-equilibrium models of black
	hole-neutron star (BHNS) binaries with mass and spin values
	compatible with parameter estimates derived from gravitational
	radiation events GW200105 and GW200115, events consistent with the
	merger of BHNSs. Using the \fuka initial-data framework, we determine
	the location of ISCO (Innermost Stable Circular Orbit) and radius of
	mass-shedding. In most of the cases studied here the innermost stable
	orbit is located at larger separations. This is consistent with the
	fact that for those two events no electromagnetic counterparts have
	been observed since it is believed that the NS will enter into the
	'plunge' phase in a short time once the separation of the components
	of the binary will be smaller of ISCO.  In analogy with classical
	binaries, we have associated to these QE sequences a Newtonian and
	Post-Newtonian Roche Lobe analysis to verify whether the NS is
	filling its Roche Lobe before approaching the ISCO. For selected
	configurations explored here, the location of ISCO and of the orbit
	at which mass shedding occurs are at separation smaller than the last
	converged solution of our sequences.  Our analysis shows that in such
	cases the neutron star is filling its Roche lobe suggesting that mass
	transfer might occur well before encountering the last stable orbit,
	and this should happen in a catastrophic way in order to prevent any
	electromagnetic emissions.  If this is the case, we are suggestion a
	third fate of the neutron star beyond the plunge or tidal disruption
	ones.
	\end{abstract}
	\maketitle
	\newpage
	\pagenumbering{arabic}

\section{Introduction}

Among the numerous gravitational wave events discovered by the LIGO,
Virgo and KAGRA collaboration, GW200105 and GW200115 events \cite{GW}, \cite{GW1}
have the strongest correlation to be the result of the merge of binaries
composed by a black hole (BH) and a neutron star (NS) usually referred as
BHNS system. The binary GW200105 has component masses  $
8.9^{+1.2}_{-1.5} M_\odot$ and $1.9^{+0.3}_{-0.2} M\odot$.  The masses
are consistent with a black hole for the primary and a neutron star for
the secondary. The primary spin is less than $0.23$ with $90\%$
confidence level, but its orientation is unconstrained. GW200115 consists
of $5.7^{+1.8}_{-2.1} M\odot$  black hole and $1.5^{+0.7}_{-0.3} M\odot$
neutron star.  The primary spin has a negative spin projection onto the
orbital angular momentum at $88\%$ probability. The spin or tidal
deformation of the secondary component for either events are not
constrained by the data. The interesting aspect of both events is that no
electromagnetic counterparts were detected.

Another event observed in 2023 May is GW230529 which is believed to have
components masses of $3.6^{0.8}_{-1.2} M_\odot$ and $1.4^{0.6}_{0.2}
M_\odot$ at the $90\%$ confidence level.  In this case, it is not
possible  to determine in definite way the nature of the primary, which,
most probable, is a black hole with mass between $2.5$ to $4.5 M_\odot$,
that resides in the lower mass gap ($3 M_\odot \lesssim m_1 \lesssim 5
M_\odot$). The spin components are not well determined, but they are
considered to be smaller than the max values relative to each components.
Also in this case no electromagnetic counterparts were detected.

The current theoretical understanding (see \cite{BHNS_review} and
references therein) is that for such parameters there is no tidal
disruption which forms an accretion disc and therefore no electromagnetic
emission occurs.

The purpose of this paper is to make a systematic study of  the
parameters space, compatible with GW200105 and GW200115, of
quasi-equilibrium sequences of BHNS binaries to explore  the properties
of their space-time without any symmetry and in a full general
relativistic regime. We determine the location  of the Innermost Stable
Circular Orbit (ISCO) for each sequence and the onset of tidal disruption
of the Neutron stars by using the Newtonian  approximated estimate given
by \cite{BHNS_review} or by the estimate of the Roche Lobe radius of the
NS as described in \cite{Ferrari}. Also,  we also discuss whether for
those systems the conditions for a dynamical instability, like the one
described in the papers  \cite{ACN},  \cite{lanza}, \cite{nishida}, \cite{nishida_lanza}, \citep{AKL},\cite{masuda},\cite{Korobkin}  named 'Runaway
instabilty', are satisfied. Such an instability, discovered for a
self-gravitating disk around a black hole,  arises when the disk overflow
its Roche lobe which exhibits a cusp at the $L_1$ Lagrange point.   In
analogy with classical close binary systems this could happen also for
the systems considered here.  According to such a picture, if the neutron
star overflows its Roche lobe then the mass transfer through the cusp will
push the cusp outwards, making a larger fraction of the disc matter
unstable to accretion. This drives the cusp out even further, and leads
to an exponential growth of the mass-transfer rate.  If such conditions
are satisfied, tidal disruption of the NS will not leave a disk around
the black hole. However, in order to test the onset of such instability,
it is necessary to do time evolution of the system. Here, we estimate the
radius of the Roche lobe and compare it with the radius of the Neutron
star.

A different scenario, proposed originally for NS-NS systems (see
\cite{stripping} and references therein), has been  recently considered
by \cite{stripping2}. According to this suggestion, as the NS approaches
the more massive primary it will fill its Roche Lobe, will lose mass and
it will move away from the primary since the asymmetry in mass will
increase. However the NS will continue to lose mass until for a
hydrodynamic instability the star will explode producing an EM burst.

We construct the quasi-equilibrium sequences by using \fuka (Frankfurt
University/KAdath Initial Data) framework \cite{FUKA}, a fork of the
\kadath spectral solver library developed by Philippe Grandclément
\cite{KADATH}). Recently, in a couple of papers \cite{topolski1},
\cite{topolski2} such framework has been used to the study of the
properties of quasi-equilibrium sequences of black hole-neutron stars
binaries and their evolution. These studies have, for the first time,
investigated the cases of black hole with high spin and large mass
asymmetries and determined an analytic formula for the mass shedding
frequency as function of mass ratio, stellar compactness and ISCO
frequency.

For each initial data sequence with fixed mass and spin ratio,  in the
present paper we determine the location of the ISCO as a function of the
separation between the two objects. It is then interesting to see how
this location will change when we change the mass  and the spin ratio.
These are then compared with the estimates (\cite{BHNS_review}) of the
orbit at which mass shedding occurs due to the black hole tidal force.

In a following paper we will describe the full evolution of such initial
data by using the Einstein Toolkit.

\section{Brief description of \fuka framework}

The results from this work are made possible using \fuka, a collection of
ID solvers to compute solutions for black holes (BBH), binary neutron
stars (BNS) and black hole neutron stars binaries (BHNS), see
\cite{FUKA}. \fuka utilizes the \textit{eXtended Conformal Thin-Sandwich}
(XCTS) system of elliptic constraint equations to compute the initial
data hypersurface.  A key feature of \fuka is its ability to reliably
compute solutions across the space of mass ratio, component spins, and
equation of state (EOS), which are critical to this study.

Here we briefly describe the numerical implementation and relevant system
of equations which are described in detail in~\cite{FUKA}.

\subsection{\textit{eXtended Conformal Thin-Sandwich}}

To construct initial data to be evolved in full general relativistic
space-time, it is necessary to start with the $3+1$ split of space-time
in slices of spacelike three-dimensional surfaces
(\textit{hypersurfaces}), $\Sigma_t$. Following this procedure the
Einstein field equations reduce to a well-posed Cauchy problem
(hyperbolic set of first order in time, partial differential equations)
subject to the constraint equations (elliptical set of second order in
space, partial differential equations) to be solved on the
\textit{hypersurfaces} at each time step.  The initial data is provided
by the solutions of the elliptical set of equations.

Many papers have been devoted to the 3+1 formalism since the beginning of
last century. For an extensive historical list of references see the
lecture notes by Eric Gourgoulhon~\cite{Eric} and also the Numerical
Relativity books by T.W. Baumgarte and S.L. Shapiro (\cite{NumRel1},
\cite{NumRel2}).

The Hamiltonian and momentum constraint equations, as the result of
projecting Einstein equations along the normal of $\Sigma_t$, are the
following:

\begin{align}
R + K^2 - K_{ij}K^{ij} = &16 \pi E,\\
D_jK^j_i - D_iK  =  &8\pi j_i,
\end{align}
where $K_{ij}$ is the extrinsic curvature of $\Sigma_t$, $E$ and $j_i$
the time and space projections of the energy-momentum tensor $T_{\mu\nu}$
respectively and $D_i$ represents the spatial covariant derivative..

We restrict our analysis to a space metric that is conformally flat:
\begin{equation}
\gamma_{ij} = \Psi^4 \tilde\gamma_{ij}\,,
\end{equation}
where $\Psi$ is a scalar field and $\tilde\gamma_{ij}$ the flat
background metric. This method is commonly referred to as the
\textit{Conformal Thin-Sandwich} decomposition or simply \textit{CTS}.
To enforce maximal slicing, $K := 0$ which implies
\begin{equation}
K_{ij}=\Psi^{-2}\hat{A} +\dfrac{1}{3}K\gamma_{ij}\,.
\end{equation}

To compute quasi-equilibrium initial data, an approximate time symmetry
must be found.  In \fuka, this symmetry is approximated using a helical Killing vector
$\xi^\mu$ given by
\begin{equation}
\xi^\mu = t^\mu = \alpha n^\mu + \beta^\mu,
\end{equation}
in a coordinate system where $t$ is the  coordinate time, $ \alpha $ the
lapse  and $ \beta $ the shift function.  A frame co-rotating with the binary
is modeled as
\begin{equation}
\beta^i_{cor} = \xi^i = \Omega\partial^i_\varphi(\textbf{x}_c) \,,
\end{equation}
so that the system can be considered quasi-stationary. Here $\Omega$ is
the orbital velocity and $\partial_\varphi^i(\textbf{x}_c)$ is azimuthal
rotation field centered about the center-of-mass of the binary. Imposing
the above approximations reduces the XCTS system to
\begin{align}
\tilde D^2\Psi & =  -\dfrac{1}{8} \Psi^{-7}\hat{A}_{ij} \hat{A}^{ij} - 2\pi \Psi^5 E, \label{eq:Hamconstraint} \\
\tilde D^2(\alpha \Psi) & =  \dfrac{7}{8}\alpha\Psi^{-7}\hat{A}_{ij} \hat{A}^{ij}+2\pi\alpha\Psi^5(E+2S), \label{eq:Lapseconstraint} \\
\tilde D^2\beta^i & =  -\dfrac{1}{3}\tilde D^i \tilde D_j \beta^j + 2 \hat{A}^{ij}\tilde D_j\left(\alpha\Psi^{-6}\right) + 16\pi\alpha\Psi^4 j^i \label{eq:Momconstraint}
\end{align}
where $E, S$ and $j^i$ are projections of the energy-momentum tensor
$T^{\mu\nu}$ (see sec. 2.2). This is a system of five equations: one for
the conformal factor $\Psi$, one for the lapse function $\alpha$ and
three for the shift vector $\beta^i$.

\subsection{Matter sources and hydrostatic equilibrium}

The neutron star matter is described by a perfect fluid, therefore the
energy-momentum tensor $T^{\mu\nu}$ is given by:
\begin{equation}
T^{\mu\nu} = (e +p) u^\mu u^\nu + p g^{\mu\nu}\,,
\end{equation}
where $e=\rho(1+\epsilon)$ is the total energy density,  $\rho$ is the
rest-mass density, $\epsilon$ the specific internal energy, $p$ the
pressure, and $u^\mu$ the fluid four velocity. The source terms entering
Eqs.~\eqref{eq:Hamconstraint} - \eqref{eq:Momconstraint} are then:
\begin{align}
E & =\rho h W^2,\\
S^i_{\hspace{1mm}i} & = 3p +(E+p)U^2,\\
j^i & = \rho h W^2 U^i,
\end{align}
where $S^{ij}$ is the fully spatial projection of the energy-momentum
tensor $T^{\mu\nu}$, $h=1+\epsilon+p/\rho$ is the relativistic specific
enthalpy, and $U^i$ the spatial projection of the fluid four velocity.
Therefore the Lorenz factor $W$ is defined as:
\begin{equation}
W^2 = (1-U^2)^{-1}.
\end{equation}
To enforce hydrostatic equilibrium conditions requires satisfying the
conservation of energy-momentum and rest mass
\begin{align}
\nabla_\mu T^{\mu\nu} &=0,\\
\nabla (\rho u^\mu) & =0.
\end{align}
To satisfy this system, \fuka assumes an isentropic unperturbed neutron
star.  Moreover, \fuka distinguishes between a co-rotating and
irrotational binary to simplify conservative equations. In the first
case, the co-rotating fluid velocity, $V^i$ is defined as
\begin{equation}
V^i =\alpha U^i -\xi^i
\end{equation}
and it is zero $V^i=0$. In the second case, it introduces a velocity
potential $\phi$ such that the enthalpy current $\hat{u}_i :=
\gamma^\mu_{\hspace{1mm}i} u_\mu = D_i\phi$ (see \cite{velpot1} and
\cite{velpot2}). For the case of an arbitrary spin, \fuka utilizes the
constant rotation velocity (CRV) approximation to model the enthalpy
current as
\begin{align}
	\hat{u}_i := D_i \phi + \omega \xi_{\rm NS}^i
\end{align}
resulting in following the hydrostatic equilibrium and momentum constraints:
\begin{align}
h \alpha W^{-1}+ \tilde{D}_i \phi V^i &= 0\,,\\
\Psi^6 W V^i \tilde{D}_i H
      + \frac{dH}{d{\rm ln}\rho}\tilde{D}_i (\Psi^6 W V^i) &= 0 \,. \label{equ:conf_3mom}
\end{align}

To close the hydrodynamic system, it is necessary to specify an EOS and
appropriate boundary conditions.  \fuka allows the use of analytic equation
of state, like single polytropic or piece-wise polytropes, but also cold
tabulated equations of state. In our calculation we used the TNTYST EOS
\cite{togashi}.

	\subsection{Boundary conditions}

In order to solve the system of Eqs, (6)-(8) together with Eqs (17) and
(18) we need to impose boundary conditions at infinity, at the horizon of
the black hole and at the surface of the neutron star.
\subsubsection{Asymptotic flatness}
Given \kadath utilizes a multi-domain approach with the outer
domain utilizing a compactified coordinate system such that spatial
infinity is mapped to a finite value, we can accurately impose asymptotic
boundary conditions in a consistent manner.  In this way, asymptotic
flatness boundary conditions are enforce on the lapse, conformal factor
and inertial shift
\begin{align}
	\alpha &\rightarrow 1,\\
	\Psi &\rightarrow 1,\\
	\beta^i_{inertial} &\rightarrow 0 ,
\end{align}
see \cite{FUKA} for the detailed discussion.
Finally, the center-of-mass is fixed by enforcing force-balance
conditions at the stellar center and that the total ADM linear momentum
measured at the asymptotic limit is minimized:
\begin{align}
	\tilde{D}_x H |_{x_{c\,,{\rm NS}}} &= 0 \,,\\
	P^i_{ADM} = \frac{1}{8\pi}\int_{S_\infty} \hat{A}^{ij} ds_j &= 0 \,.
\end{align}

	\subsubsection{Black-hole excision boundary conditions}
In numerical simulations the presence of a black hole introduces a
singularity unless suitable boundary conditions are enforced. For initial
data, it has become common practice to excise the surface of the black to
avoid this. The excision region for conformally flat initial data is
defined as the marginally outer trapped surface (MOTS) such that the
vector field $k_\mu$ of outgoing null rays on the surface vanishes on
them (see \cite{excision1}, \cite{excision2}).  This naturally imposes
the following boundary conditions on the excision surface:
\begin{align}
	\beta^i|_{\mathcal{S}\,, {\rm BH}} &= \alpha \psi^{-2} \tilde{s}^i + \omega \xi^i_{\rm BH} \,,\\
	\tilde{s}^i \tilde{D}_i \left(\alpha \psi\right)|_{\mathcal{S}\,, {\rm BH}} &= 0 \,,\\
	\tilde{s}^i \tilde{D}_i \psi|_{\mathcal{S}\,, {\rm BH}} &=
		- \frac{\psi}{4} \tilde{D}_i \tilde{s}^i
		- \frac{1}{4} \psi^{-3} \hat{A}_{ij} \tilde{s}^j \tilde{s}^i \,,
\end{align}
where $\tilde{s}^i$ is the conformal outward pointing normal vector to
the excision surface and $\xi^i_{\rm BH}$ is the rotational vector field
center on the black hole, with rotation velocity $\omega$.

	\subsubsection{On the surface of the neutron star}

The surface of the neutron star is defined as that surface in which $H :=
\ln\left(h\right) \rightarrow 0$. From the equation of hydrostatic
equilibrium Eqs.~\eqref{equ:conf_3mom} the boundary condition of the
surface of the neutron star becomes:
\begin{equation}
	V^i \tilde D_i H=0\,.
\end{equation}

\subsection{Diagnostics}
To characterize the properties of the quasi-equilibrium sequences, we
define the relevant diagnostics here that will be used extensively in our
results. Specifically, in the asymptotic limit, one can write the total
energy and angular momentum following the ADM formalism as
\begin{align}
	M_{ADM}&=-\frac{1}{2\pi}\int_{S_\infty} D^i\Psi ds_i \,,\\
	J_{ADM}&=\hspace{2mm}\frac{1}{8\pi}\int_{S_\infty} \hat{A}^{ij} \xi_i ds_j\,.
\end{align}
Since the spacetime is stationary and it has an approximate Killing
vector, we also evaluate the Komar mass $M_K$, which is
expressed as a surface integral at infinity
\begin{equation}
	M_K = \frac{1}{4\pi} \int_{S_\infty} D^i \alpha ds_i.
\end{equation}
By using the ADM mass, the binding energy between the two objects is
calculated by taking the difference of the ADM mass of the system and the
sum of the ADM masses $M_{1,2}$ of each single component in isolation:
\begin{equation}
	E_b = M_{ADM} -(M_1+M_2)
\end{equation}

\section{\fuka numerical strategy}

In order to solve the constraint equations, \fuka \cite{FUKA} uses an
extended version of the \kadath spectral library \cite{KADATH} which make
use of a novel multi-domain discretization which maps the physical space
to an optimized numerical grid. For binary ID, a bi-spherical
decomposition is used for the numerical space resulting in at least
twelve domains to resolve the problem efficiently (see \cite{KADATH} and
\cite{FUKA} for a more detailed discussion). The equations resulting from
the spectral representation are solved by using Newton-Raphson iterative
method~\cite{KADATH}.

The libraries is modular and one has to specify which solver to use
according to the nature of the binary considered BBH, BNS or BHNS. In all
the cases, the code first finds a converged solution for each isolated
object belonging to the binary. These solutions for the isolated objects
are then combined together and the constraint equations are solved
following different stages in which specific assumptions/simplifications
are made in order to progressively obtain a good initial guess for the
successive stage. Finally, to minimize the residual drifts of the center
of mass by requiring that the linear ADM momentum are minimized:
\begin{equation}
P^i_{ADM} = 0.
\end{equation}

A characteristic of \fuka framework is that the converged solution
relative to each different stages are saved in files from which a new run
could start. This is very useful for small clusters where the wall time
of the single queue may not be long enough to get a final converged
solution. In this case one can restart the process from the last
converged stage. Also, if a high resolution solution is needed at a later
time, one can start from the converged lower resolution solution and
avoid costly recalculation of the ID solution from scratch.

The robustness and accuracy of \fuka on constructing quasi-equilibrium
sequences of BHNS binaries has been analyzed thoroughly in
\cite{topolski1}, where it has shown excellent agreement between their
sequences with those published in \cite{Taniguchi} by comparing the
mass-shedding indicator $\kappa$ (see Eq.  (\ref{eq:kappa}) for a
definition) and the binding energy $E_{b}$ along BHNS sequences of
constant mass ratios by using polytropic equation of state with
$\Gamma=2$.

\section{Numerical results}

	In this paper we  have constructed  quasi-equilibrium sequences of
	BHNS binaries whose components have masses compatible with those
	estimated from the sources GW200105 and GW200115 and using the
	Togashi realistic equation of state (\cite{togashi}. In particular,
	behind the most probable values of the masses, we have also
	constructed sequences for the allowed extreme values.  Different
	dimensionless spins $\chi$ for each component have been chosen
	according to Fig. 6 of paper \cite{GW} where the two-dimensional
	posterior probability of the spin-tilt and spin magnitude for both
	components of the binary are shown for the two sources.  The range of
	possible spins extracted from GW200105 for the BH is
	$-0.2<\chi_{BH}<0.2$ and $-0.4<\chi_{NS}<0.4$ for the NS. In the case
	of GW200115, the component dimensionless spin ranges are
	$\chi_{BH}<0.0$ and $\chi_{NS}<0.4$. However, for both events, the
	data does not allow for strong constraints on their values although
	there is high probability of negative alignment of at least one
	component.

	The detailed parameters used in the calculations presented in this
	paper are shown in Table \ref{table:GW2001X5} for both GW sources.
	In this Table,  $M_{BH}= M_{irr}$ and $M_{NS}$ is the ADM mass of the
	isolated Neutron star $M_{ADM,\infty}$.
\begin{table}[!ht]
\centering
\begin{tabular}{|c|c|c|c||c|c|c|c|}
\hline
  \multicolumn{4}{|c||}{ \textbf{GW200105 }}&\multicolumn{4}{|c|}{ \textbf{GW200115 }}  \\
  \hline
   \multicolumn{4}{|c||}{$\chi_{BH}=-0.1$ \quad $\chi_{NS}=0.3$}&\multicolumn{4}{|c|} {$\chi_{BH}=-0.2$ \quad $\chi_{NS}=0.2$} \\
  \hline
$q$&$Q$&$ M_{BH}$&$M_{NS}$ &$q$&$Q$&$ M_{BH}$&$M_{NS}$     \\
\hline
0.2973&3.3636&7.40&2.20 &   0.6111& 1.6364 & 3.60 & 2.20  \\
 \hline
0.2567&3.8947&7.40&1.90&0.4167& 2.4000 & 3.60 & 1.50   \\
 \hline
0.2472&4.0455&8.90&2.20&0.3860& 2.5909 & 5.70 & 2.20     \\
 \hline
0.2297&4.3529&7.40&1.70&0.3333& 3.0000 & 3.60 & 1.20    \\
  \hline
0.2178&4.5909&10.10&2.20&0.2933& 3.4091 & 7.50 & 2.20     \\
 \hline
\textbf{0.2135}&\textbf{4.6842}&\textbf{8.90}&\textbf{1.90}&\textbf{0.2632} & \textbf{3.800} & \textbf{5.70} & \textbf{1.50}      \\
 \hline
0.1910&5.2353&8.90&1.70&0.2105& 4.7500 & 5.70 & 1.20     \\
  \hline
0.1881&5.3158&10.10&1.90&0.2000& 5.0000 & 7.50 & 1.50    \\
 \hline
0.1683&5.9412&10.10&1.70&0.1600& 6.2500 & 7.50 & 1.20    \\
  \hline
 \end{tabular}
 \caption{List of parameters used to calculate quasi-equilibrium
 sequences for the two events GW200105 and GW200115. The parameter q is
 defined as $q:=M_{NS}/M_{BH}$ and $Q:=q^{-1}$. In bold characters are shown
 the most probable values  for the two events. }
 \label{table:GW2001X5}
 \end{table}

\begin{figure*}[!ht]
	\centering
	\includegraphics{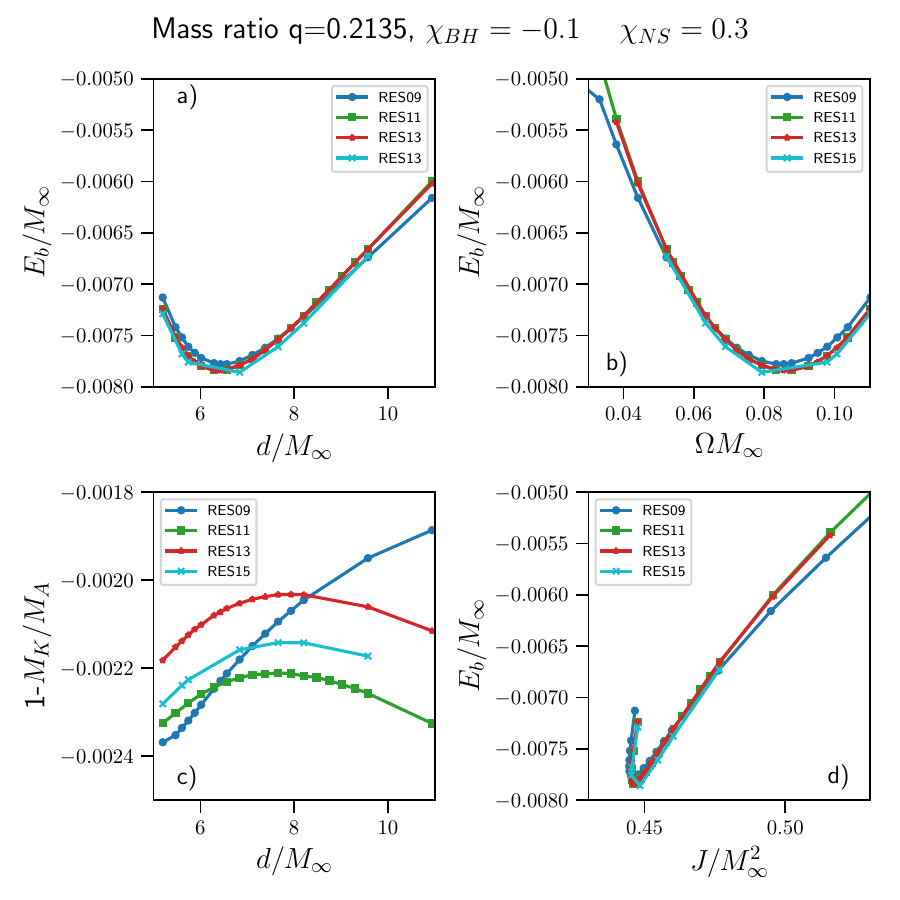}
	\caption{a) Binding energy $E_b/M_{\infty}$ versus separation $d/M$  and
	b) versus the orbital angular velocity; c) {\it virial error}
	$1-M_K/M_{ADM}$ versus separation; d) binding energy versus total angular
	momentum $J/M_{\infty}^2$. All the quantities are shown at different
	resolution RES. }
	\label{fig:conv1}
\end{figure*}
	Other sequences with different pairs of spin components have also been
	computed, here we present only the  two cases of Table
	\ref{table:GW2001X5} because they are more representative for
	determining the location of ISCO and the relative distance of the
	binary components for the onset of tidal disruption.
	According to what was found in \cite{topolski1}, we choose negative
	rotating BHs since predominantly lead to plunge configurations, in
	fact, in these cases,  the ISCO moves further outwards, therefore, it
	is challenging for the NS to be disrupted outside of the ISCO.



In \fuka, the spectral resolution is defined by the number of collocation
 points in $r$, $\theta$ and $\phi$ directions. The spectral resolution
 across all domains in \fuka is currently set by a single parameter
 (RES), such that ${\rm RES} = 9$ implies $\left(r \,, \theta \,, \phi
 \right) = \left( 9 \,, 9 \,, 8 \right)$, in each domain. Here we
 determine the convergence of the ID by examining a sequence of BHNS
 initial data with mass ratio $q=0.2135$ and component spins
 $\chi_{NS}=0.3$ and $\chi_{BH}=-0.1$, using four different resolution
 ${\rm RES} = 9 \,, 11 \,, 13 \,, 15$. In Fig.~\ref{fig:conv1} we present
 the sequence for a mass ratio $q=0.2135$ with component spins
 $\chi_{NS}=0.3$ and $\chi_{BH}=-0.1$. Overall, we find the convergence
 of the solution is already acceptable for ${\rm RES} = 11$. Therefore,
 the remainder of the results discussed in this work use ${\rm RES} = 11$
 exclusively for all sequences considered.
The plots in Fig.~\ref{fig:conv1}  also show  a clear minimum on the
binding energy.  Its  relative position corresponds to the location of
ISCO for a given binary configuration.

 \subsection{Quasi-equilibrium sequences for the GW200115 event}

In this section we analyze the numerical results obtained from computing
quasi-equilibrium sequences consistent with the GW200115 event. The
results for the GW200105 sequences are presented in
Appendix~\ref{GW200105}, which are consistent with our findings for
GW200115. The aim of our analysis is to find the ISCO radii, for a set of
parameters compatible with the observed ones, and compare them with the
estimates of the radii at which mass shedding occurs, by using different
approaches.

\begin{figure*}[t]
	\includegraphics[scale=0.45]{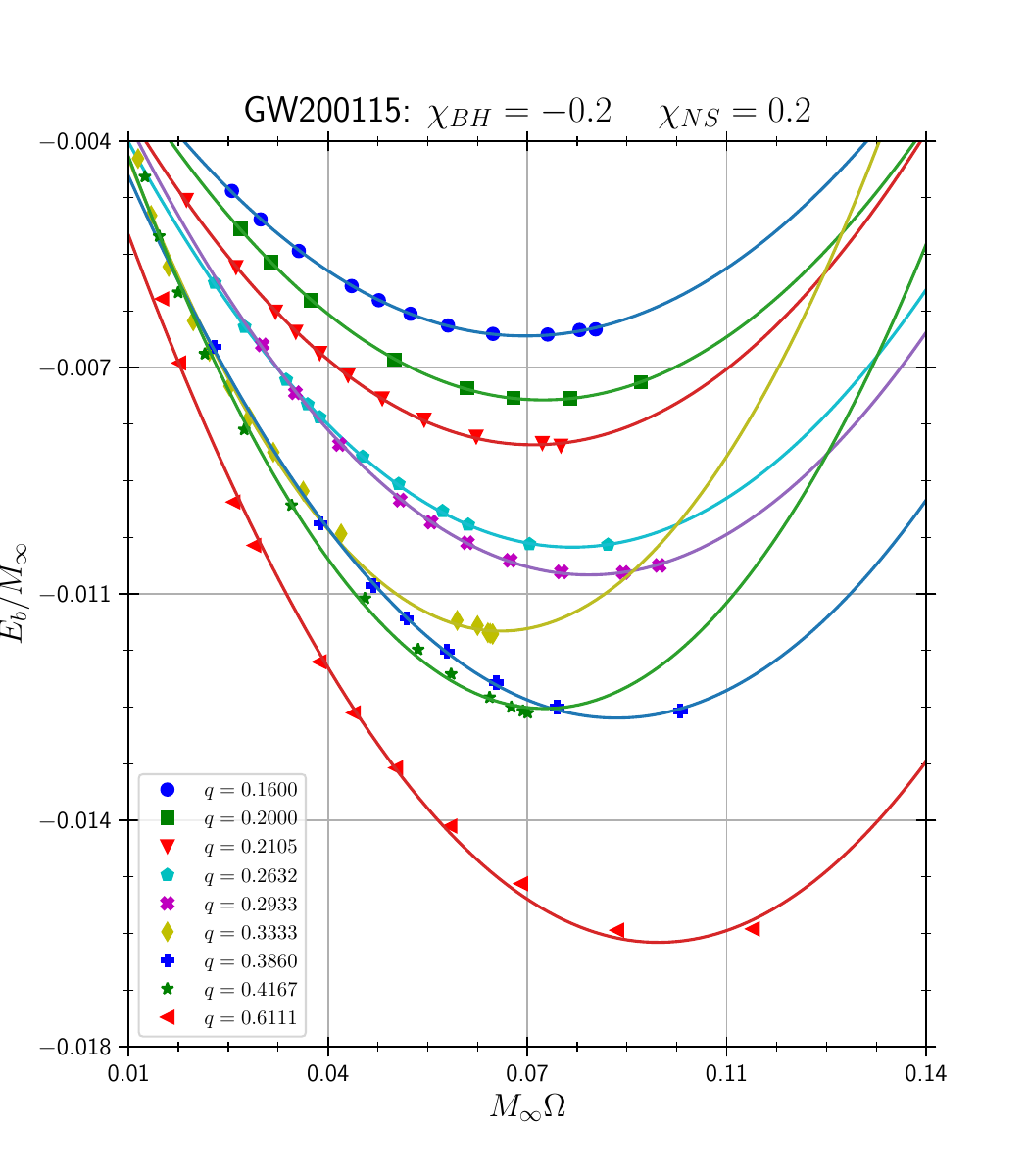}
	\includegraphics[scale=0.45]{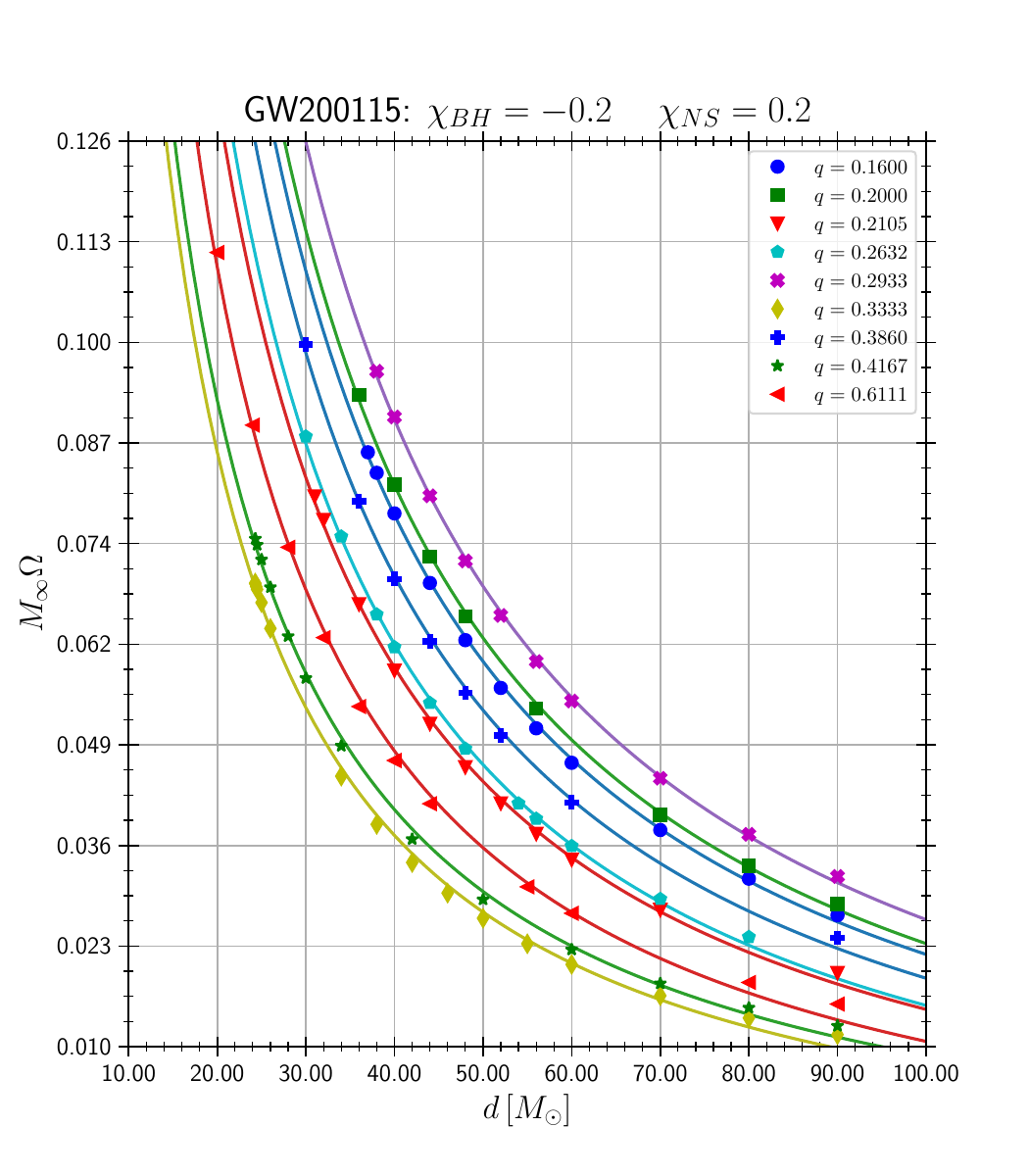}
	\caption{\textit{Left panel:} Binding energy versus orbital angular
	velocity for different mass ratios with constant spin $\chi_{NS}=0.2$
	for the neutron star and $\chi_{BH}=-0.2$ for the black hole for
	sequences simulating the GW200115 event. \textit{Right panel:
	}Orbital angular velocity versus binary separation. In both panels,
	different colored symbols represent the values coming from the
	calculated sequences, solid lines of the same color report the fit
	with the functional form given by Eq. (\ref{eq:parabola}) for
	$E_b/M_\infty$ and by Eq. (\ref{eq:hyperbola}) for
	$M_\infty\Omega_{ISCO}$.}
	\label{fig:Eb_Omega_fit_gw115}
\end{figure*}

\subsection{ISCO location}

To find the location of the innermost stable circular orbit, we need to
find the minimum of the binding energy of the binary. The values of
minima for the binding energy can be found by first interpolating the
sequence by using cubic spline and then applying some method of
minimization to isolate the minimum (see for instance \cite{NumRec} and
reference therein).

However, in some  cases the code does not converge for smaller
separations and it is not possible to find a minimum of the binding
energy.  Following the suggestion in \cite{topolski1},  we use the
following functional ansatz to fit the computed data and extrapolate to
find the minimum
\begin{equation}
 E_{b,fit}= b_1 +b_2 \Omega + b_3 \Omega{^2} \,, \label{eq:parabola}
\end{equation}
where $b_{i}| i=\left\lbrace 1\,, 2\,, 3\right\rbrace$ are the parameters
to be determined by the fitting process. The minimum of the fitting
function will then determine the orbital angular velocity corresponding
to the ISCO
\begin{equation}
 \frac{\partial E_{b,fit}}{\partial \Omega}\bigg|_{\Omega_{ISCO}} =0 \,. \label{equ:Omega_ISCO}
\end{equation}
The radius of the marginally stable orbit can then be determine by
fitting the orbital angular velocity with the ansatz
\begin{equation}
M_\infty\Omega = \frac{a}{d} +b \,,\label{eq:hyperbola}
\end{equation}
where the parameters $a$ and $b$ depends on the sequence considered. The
value $d_{ISCO}$ is then found once we insert $M_\infty\Omega_{ISCO}$
obtained from Eqs.(\ref{eq:parabola})-(\ref{equ:Omega_ISCO}) into
Eq.(\ref{eq:hyperbola}). The data and the resulting fits are illustrated
in Fig.~\ref{fig:Eb_Omega_fit_gw115}.

\begin{table*}[!t]
    \centering
    \begin{tabular}{|c|c|c|c|c|c|c|c|}
    \hline
  \multicolumn{8}{|c|}{ \textbf{GW200115 }}  \\
  \hline
   \multicolumn{8}{|c|}{$\chi_{BH}=-0.2$ \quad $\chi_{NS}=0.2$} \\
 \hline
         q & b1 & b2 & b3 & $M_{\infty}\Omega_{ISCO}$ & a & b & $d_{ISCO} [M_\odot]$ \\ \hline
        0.6111 & -0.00278 & -0.2825 & 1.4661 & 0.0963 & 2.48 & -0.0141 & 22.45 \\
        0.4167 & -0.00155 & -0.2886 & 1.8542 & 0.0778 & 2.10 & -0.0121 & 23.34 \\
        0.3860 & -0.00231 & -0.2369 & 1.3230 & 0.0896 & 3.43 & -0.0156 & 32.66\\
        0.3333 & -0.00167 & -0.2805 & 1.9856 & 0.0706 & 1.96 & -0.0120 & 23.76 \\
        0.2933 & -0.00174 & -0.2109 & 1.2401 & 0.0850 & 4.27 & -0.0164 & 42.08 \\
        0.2632 & -0.00217 & -0.1969 & 1.1960 & 0.0823 & 3.08 & -0.0156 & 31.51 \\
        0.2105 & -0.00191 & -0.1786 & 1.1761 & 0.0759 & 2.92 & -0.0144 & 32.31 \\
        0.2000 & -0.00148 & -0.1682 & 1.0851 & 0.0775 & 3.91 & -0.0160 & 41.89 \\
        0.1600 & -0.00159 & -0.1452 & 0.9719 & 0.0747 & 3.75 & -0.0157 & 41.52 \\ \hline
    \end{tabular}
    \caption{Fitting parameters for the binding energy $Eb/M_{\infty}$
    ($b_1, b_2, b_3$) and the value of the orbital angular velocity at
    its minimum $M_\infty\Omega_{ISCO}$.  The table also shows the
    fitting parameters for the orbital angular velocity $M_\infty\Omega$
    as a function of binary separation along with the computed binary
    separation distance ($d_{ISCO}$) of the ISCO.}
    \label{table: fit_params_GW115}
\end{table*}

In Table \ref{table: fit_params_GW115} we show the fitting parameters
$b_i$ for $i=\left\lbrace 1 \,, 2 \,,3 \right\rbrace$ and the values of
$M_\infty\Omega_{ISCO}$. In addition, the fitting parameters $a$ and
$b$ along with the values of $d_{ISCO}$ are shown for each sequence
simulating the event GW200115.
The values of $d_{ISCO}$ should now be compared with the estimate of
the separations at which mass shedding occurs for each sequence
calculated, which will will now explore in the next section.

\subsection{Mass shedding radius location}
To precisely determine the separation of the two components corresponding
to the onset of tidal disruption, it is necessary to compute the full
time evolution of the system starting from an initial data solution such
as those discussed here. This is the topic of a successive paper.
However, we can use several suggestions present in the literature to
estimate it and make a comparison with the values of ISCO found in our
sequences.

In~\cite{BHNS_review}, the authors suggest that tidal disruption of the
secondary will occur only after substantial mass is stripped from the
surface of the neutron star due to mass shedding induced by the black
hole when its gravity overcomes the self-gravity of the neutron star at
its inner edge. By using Newtonian arguments, they determine the
mass-shedding limit as the orbit with radius given by:

\begin{equation}\label{eq:rMS}
r_{MS} = 2^{1/3} c_R \left( \frac{M_{BH}}{M_{NS}}\right)^{1/3} R_{NS}
\end{equation}
where $c_R\geq 1$ represents the degree of elongation of the stellar
radius due to tidal force. We estimate  $c_{R}$ taking the inverse of the
mass-shedding indicator $\kappa$, introduced in \cite{Taniguchi},
calculated at the end of each sequence by the relation:
\begin{equation} \label{eq:kappa}
\kappa\equiv \frac{\left(\partial(\ln h)/\partial r\right)_{eq}}{\left(\partial(\ln h)/\partial r\right)_{pole}}.
\end{equation}
The indicator goes from one, for a spherical star, to zero when a cusp is
formed at the surface. Table~\ref{table:tidal_115} includes the values of
$c_R:=1/\kappa$, the location of the mass-shedding radius and the ISCO for
the sequences relative to GW200115.

 \begin{table}[!ht]
 
\centering
\begin{tabular}{|c|c|c|c|c|c|}
\hline
  \multicolumn{6}{|c|}{ \textbf{GW200115 }}  \\
  \hline
   \multicolumn{6}{|c|}{$\chi_{BH}=-0.2$ \quad $\chi_{NS}=0.2$} \\
  \hline
$q$&$d_{_{\rm min}} [M_\odot]$&$R_{NS} [M_\odot]$&$r_{ISCO} [M_\odot]$&$r_{MS} [M_\odot]$&$1/\mathcal{\kappa}$   \\
\hline
0.6111&18.00&7.27&22.45&14.34&1.33     \\
 \hline
0.4167&24.30&7.76&23.34&19.11&1.46   \\
 \hline
0.3860&26.00&7.24&32.66&14.81&1.18     \\
 \hline
0.3333&24.28&7.74&23.76&28.88&2.05 \\
  \hline
0.2933&40.00&7.25&42.08&15.46& 1.13  \\
 \hline
0.2632&30.00&7.67&31.51&20.15& 1.33    \\
 \hline
0.2105&31.00&7.66&32.31&25.02& 1.54  \\
  \hline
0.2000&36.00&7.67&41.89&20.65& 1.25  \\
 \hline
0.1600&37.00&7.63&41.52&22.42&1.27   \\
  \hline
 \end{tabular}
  \caption{We provide here $d_{_{\rm min}}$, the minimum separation along
  the sequence for a given mass ratio. At $d_{_{\rm min}}$ we compute
  $R_{NS}$, the proper circumferential radius of the NS as measured on
  the stellar surface; $r_{MS}$, the estimated mass shedding radius given
  by Eq.~\eqref{eq:rMS}; and $1/\kappa$, given by Eq.~\eqref{eq:kappa}.}
 \label{table:tidal_115}
 \end{table}

The use of $\kappa$ as a mass-shedding indicator has been explored for QE
sequences originally in \cite{Taniguchi} and more rigorously in
\cite{topolski1}, namely to extract the critical orbital velocity
$\Omega_{MS}$, the estimated orbital velocity when the neutron star
starts to shed mass. While this has been shown to be a useful tool to extract
information related the occurrence of mass-shedding, it still relies on a chosen
model and extrapolation to determine $\Omega_{MS}$.  In the following section,
we will explore a novel approach based on (post-)Newtonian analysis to provide
physical insights into the mass-shedding radius and the Roche lobe of the
neutron star.

 \subsection{Newtonian equipotential analysis of QE solutions}

A different approach to determine $r_{MS}$ was described by
\textit{Ferrari et al}~\cite{Ferrari} and is based on the estimate of the
Roche lobe radius of the Neutron star. In~\cite{Ferrari}, they calculate
the maximum of the Newtonian potential $U(x,y)$ for a particle with mass
$m_0\ll M_{NS}\leq M_{BH}$ for equatorial orbits (see also \cite{Frank}
for a discussion on the Reduced Newtonian Effective Potential for three
body system):
\begin{equation} \label{eq:roche}
	U(x,y) = -\frac{GM_{NS}}{\sqrt{(x-r_{NS})^2+y^2}} -\frac{GM_{BH}}{\sqrt{(x-r_{BH})^2+y^2}} -
	\frac{1}{2}\omega^2 r^2 \,,
\end{equation}
where $\textbf{r}_{NS}$ and $\textbf{r}_{BH}$ are the position vectors,
with respect to the center of mass, of the center of neutron star and
Black hole respectively. The last term is the centrifugal contribution
where $\omega$ is given by

\begin{equation}
	\omega = \sqrt{\frac{G(M_{NS}+M_{BH})}{r_{12}^3}} \,,
\end{equation}
and $r_{12}$ is the radial distance of the two objects. Within this
model, mass shedding is defined to be the moment when the surface of the
neutron star touches the Roche lobe.  Crossing of the NS's Roche lobe is
a critical point, a cusp, where the gravitational forces between the two
objects balance out such that mass transfer between them is possible.
This point corresponds to the location of the inner Lagrangian point
(L1).

Such analysis has been generalized previously for General Relativistic
configurations in the presence of some symmetry. In particular, for a
system made by a self gravitating disk rotating, with constant specific
angular momentum, around a Black hole in an axisymmetric spacetime, the
Roche lobe analysis has allowed to hypothesize that a dynamical
instability with millisecond time scales (runaway instability) might
occur when the disk overflows its Roche lobe (see \cite{ACN}-
\cite{Korobkin}).

\begin{figure*}[!t]
	\centering
	{\includegraphics[width=.95\linewidth]{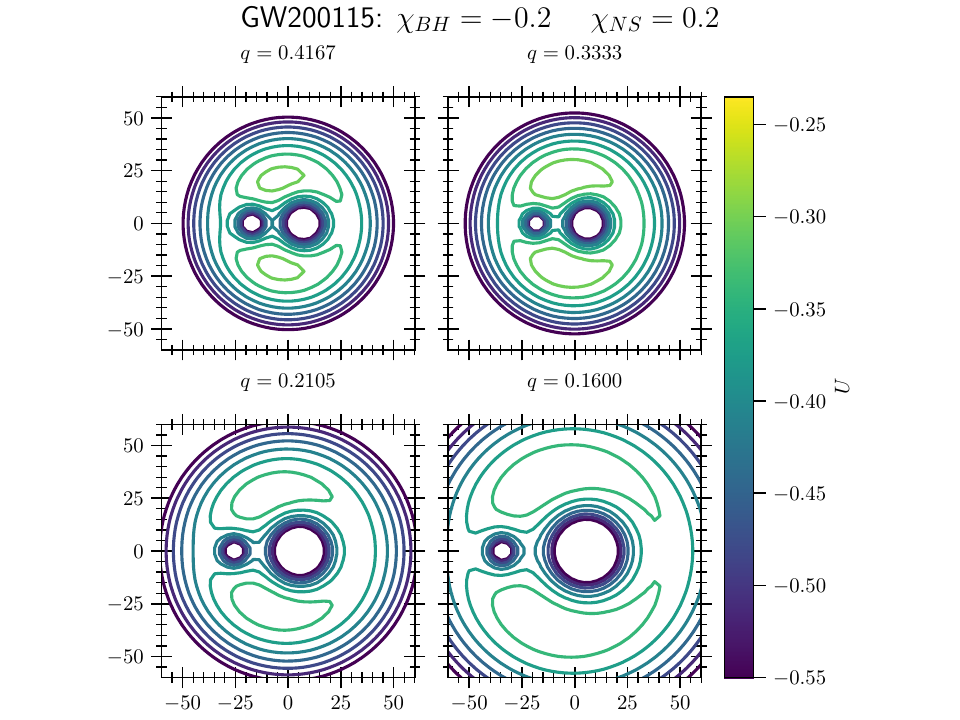}}
	\caption{Reduced Newtonian Effective Potential surfaces for four  QE
	sequences with different mass ratios  in a frame with origin at the
	center of mass. The plots refer to the last converged solution of each
	sequence to simulate GW200115 with spin $\chi_{NS}=0.2$ and
	$\chi_{BH}=-0.2$. }
	 \label{fig:equi115}
	\end{figure*}

In analogy with the physics of classical binaries whose evolution can be
understood by using Roche Lobe analysis, we have calculated the
equipotential surfaces Eq.~(\ref{eq:roche}), on the $x-y$ plane using the
values of the required parameters given by our QE sequences. In
Fig.~\ref{fig:equi115} we show the Newtonian equipotential surfaces for
the last converged solution of QE equilibrium sequences for GW200115 with
different mass ratios. For brevity, we restrict our analysis to $q =
\left\lbrace 0.4167 \,, 0.3333 \,, 0.2105 \,, 0.1600 \right\rbrace$.

In the case of mass ratio q=0.4167 we see an equipotential surface having an eight-shape which is the characteristic of a Roche lobe.

To complete the picture of the analogous Newtonian systems, we computed
the position of the three collinear Lagrange points $L_1$, $L_2$ and
$L_3$ along the $x$ axis, that is those points where no force will be
exerted and any particle there will remain at rest. Since the stars in
our binaries have comparable mass, in order to find  their location it is
necessary to find the solution of the following cubic equation (see for
instance \cite{restricted} and references therein for a derivation of
it):
\begin{equation}
	x-\frac{1-\alpha}{|x+\alpha|^3}(x+\alpha)-\frac{\alpha}{|x-1+\alpha|^3}(x-1+\alpha) = 0 \,,
\end{equation}
where $\alpha=\frac{M_{NS}}{M_{BH}+M_{NS}}$. The equation has been solved
numerically by using the Newton method.

In Table \ref{table:q04167} we report the value of the NS radius $R_{NS}$ to be compared with the Eggleton's  and Post Newtonian estimates of the radius of the Roche lobe $RL_{E}$,  $RL_{PN}$ and the position, along the x-axis, of the $L_1$ lagrangian point for different separations $d[M_{\odot}]$.   The values refer to the sequence with mass ratio $q=0.4167$.

\begin{table}[!ht]
    \centering
    \begin{tabular}{|c|c|c|c|c|c|}
    \hline
     \multicolumn{6}{|c|}{ \textbf{q=0.4167 }}\\
    \hline
        $d[M_{\odot}]$ & $R_{NS}$ & $RL_{E}$ &$RL_{P}$ & $RL_{PN}$& $L_{1}$  \\ \hline
        24.3 & 7.76 & 7.44 &7.47 &6.90 & 7.16  \\ \hline
        24.5 & 7.76 & 7.50 & 7.53 &6.96& 7.22  \\ \hline
        25.0 & 7.76 & 7.66 & 7.68 & 7.11& 7.37  \\ \hline
        28.0 & 7.77 & 8.58 & 8.60 & 8.03&8.25  \\ \hline
        30.0 & 7.77 & 9.19 & 9.22 & 8.64&8.80 \\ \hline
    \end{tabular}
    \caption{The table shows, for the sequence with mass ratio   $q=0.4167$, the NS proper radius $R_{NS}$ as given by our QE sequence, the Eggleton \cite{eggleton} and Paczynski \cite{paczynski} Roche lobe radii, $RL_{E}$ and $RL_{P}$, the Post Newtonian Roche lobe radius $RL_{PN}$ \cite{ratkovic} and the position of the $L_{1}$ lagrangian point for the smaller separations close to the end of the sequence. }
    \label{table:q04167}
\end{table}
For separations $d[M_{\odot}] \le 25$,  $R_{NS} > RL_{PN}$ and $R_{NS} > RL_{LE}$ tsuggesting that the NS star is overflowing the Roche lobe in the limit of the  (post-)Newtonian approximation used in our analysis.

\begin{figure*}[!t]
\centering
{\includegraphics[scale=0.95]{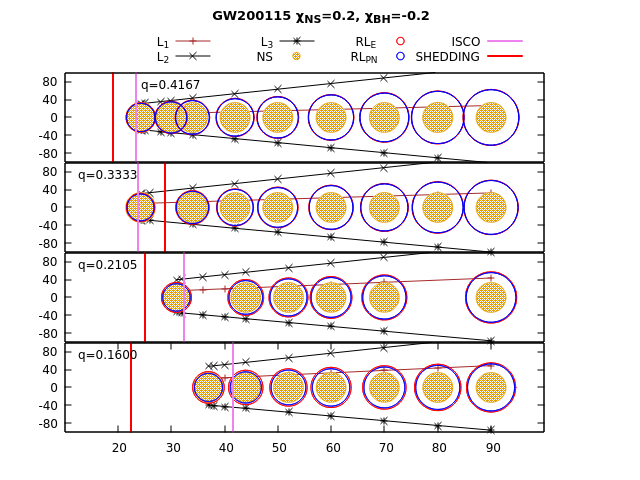}} \caption{Radii of reduced
Newtonian Effective Potential surfaces for different separation compared
with the relative positions of the collinear Lagrange points $L_1$, $L_2$
and $L_3$.  The filled circle represents the Neutron star with the proper
radius given by the QE sequence,  the red circunference is a circle whose
radius is given by the approximated Eggleton's radius of the Roche lobe
\cite{eggleton}, the radius of the blue circle has been calculated by
using the Post-Newtonian formula given in \cite{ratkovic}. Violet
vertical line corresponds to the ISCO radius and the red one to the
radius at which mass shedding starts to be important. These cases
correspond to sequences that simulate GW200115 event with spin
$\chi_{NS}=0.2$ and $\chi_{BH}=-0.2$ and different mass ratios.}
 \label{fig:radii115}
\end{figure*}

In purely Newtonian approximation, and assuming that either the Roche
lobe and the Neutron star can be both represented by a sphere, we plot in
Fig.~\ref{fig:radii115} the location of the Lagrangian points ($L_1$,
$L_2$ and $L_3$) in the $x-y$ plane together with  circles representing
the Roche Lobe with the approximate radius given by Eggleton ($RL_E$)
\cite{eggleton}, with a second Post Newtonian radius ($RL_{PN}$)
\cite{ratkovic}, and a filled yellow circle representing the Neutron star
with radius $R_{NS}$. In addition, the location of ISCO radius, as
determined by the minimum of the binding energy,  and the location at
which mass transfer will occur, computed using Eq.~\eqref{eq:rMS}, are
drawn for each configuration.
Again, we show only four representative sequences corresponding to mass
ratios $q = \left\lbrace 0.4167 \,, 0.3333 \,, 0.2105 \,, 0.1600
\right\rbrace$.

These figures show that at the end of the sequences, for small separation,
the Neutron star has the same approximated volume as that of the Roche
lobe. In the limit of the used Newtonian approximation, this suggest that
the Neutron stars is nearly filling its Roche lobe. However, in order to understand  if the NS will reach such a stage, it is necessary to compare the value of the separation of the binary with the location of ISCO. If the NS overflows its Roche lobe at separation greater than ISCO then it becomes unbounded and a process of mass transfer will start modifying the content of mass and angular momentum of
each component.

In particular, in the case of mass ratio $q=0.4167$, (see Fig.
\ref{fig:radii115}), not only is the Neutron star filling its Roche lobe,
but also the $L_1$ Lagrangian point is close to the surface of the star.
In addition, the ISCO and mass shedding radii are located at smaller
separation. In such cases, the corresponding gravitational radiation
event, without electromagnetic counterpart, should be produced by a
process different than a 'plunge'. Such conditions sound very similar to
the scenario of the unset of a runaway instability, but this could be
verified only by doing a full time evolution of such initial conditions.

For the mass ratio $q=0.3333$ in Fig.~\ref{fig:equi115}, the location of
mass shedding occurs at larger separation with respect to ISCO location.
In this case,  the NS should
be destroyed by tidal forces forming a disk around the black hole giving
rise to an electromagnetic counterpart. If not, then probably when the
Neutron star loses mass,  the mass ratio will change and the binary may
follow a different QE sequence with a different ISCO radii. 

We should emphasize though that in the cases discussed above, the mass of
the black hole is close to the lower mass gap whose existence is
uncertain although the Ligo/Virgo collaboration has reported a few
candidate events with component mass falling in the gap.

\section{Conclusions}

In this paper we have constructed a series of QE sequences of BHNS
binaries by using values of masses and spins of the components compatible
with those measured in the events GW200105 and GW200115. These values are
also consistent with those observed in GW230529. In this study we have
used Togashi equation of state \cite{togashi} which belongs to the
category of stiff EOS.

After testing the convergence of \fuka, by using different resolutions,
for each sequence we have calculated the location of the radius of ISCO
through the minimum of the binding energy $E_b$. We estimate the location
of the radius of mass-shedding by  Eq. (\ref{eq:rMS}), following
\cite{BHNS_review}, and using as degree of elongation the inverse of
$\kappa$ as defined in Eq. (\ref{eq:kappa}) given by our QE solutions. Although
this estimate is based  on Newtonian arguments it is sufficient to make
comparison with the location of ISCO. All the different estimates show
that the mass shed limit occurs mainly at separations smaller than the
last quasi-equilibrium converged solution for each sequence considered,
whereas ISCO occurs at larger radii with the exception of the case with
$q=0.3333$ where ISCO is located at a smaller radius with respect to the
mass shedding radius, see Table \ref{table:summary}.
Excluding this last case, the results are consistent with the absence of electromagnetic counterparts for the events considered here since it is believed that the NS will enter into the 'plunge' phase. 

We also made an analysis of the last QE configurations in  the framework
of Newtonian Roche Lobe used for classical binary system to understand
the fate of the secondary component.  This has been done by calculating
the radius of the  Roche lobes and relative Lagrangiam points. In particular, for the mass ratio $q=0.4167$, we found that the secondary NS
star might fill its Roche lobe before reaching the ISCO since the proper radius of the NS,  $R_{NS}$, is greater than the radius of its Roche lobe given by  the Post-Newtonian approximation \cite{ratkovic} and by Eggleton's formulae \cite{eggleton}, see Table  \ref{table:summary}. 
In such a case, since the NS is filling its Roche lobe, mass transfer  start to be important before encountering the last stable orbit. To prevent any electromagnetic emissions, the transfer should occur catastrophically in a similar way as would happen when a 'runaway instabilities' sets in.
However, in order to verify the onset of
such instability, it is necessary to do time evolution of the system by using as initial data the QE solutions described in the paper.

\begin{table*}[!ht]
\footnotesize
\centering
\begin{tabular}{|c|c|c|c|c|c|c||c|c|c|c|c|c|c|}
\hline
  \multicolumn{7}{|c||}{ \textbf{GW200105 }}&\multicolumn{7}{|c|}{ \textbf{GW200115 }}  \\
  \hline
   \multicolumn{7}{|c||}{$\chi_{BH}=-0.1$ \quad $\chi_{NS}=0.3$}&\multicolumn{7}{|c|} {$\chi_{BH}=-0.2$ \quad $\chi_{NS}=0.2$} \\
  \hline
$q$&$d[M_\odot]$&$r_{ISCO}$&$r_{MS}$&$R_{NS}$&$RL_E$ &$RL_{PN}$&$q$&$d[M_\odot]$&$r_{ISCO}$&$r_{MS}$&$R_{NS}$&$RL_E$ &$RL_{PN}$   \\
\hline
0.2973&32.00&39.32&17.62& 7.34&8.97&7.46&
0.6111& 18.00&22.45&14.34&7.27 &6.07&5.90 \\
 \hline
0.2567& 38.00&39.19&20.16&7.67 &10.24&8.61&
0.4167&24.30 &23.34&19.11& 7.76&7.44&6.90 \\
 \hline
0.2472& 38.00&46.91&18.71& 7.33&10.14&8.14&
0.3860&26.00 &32.66&14.81& 7.24&7.81&6.86    \\
 \hline
0.2297&36.00 &39.02&21.38& 7.72&9.41&7.71&
0.3333& 24.28&23.76&28.88&7.74 &7.02&6.33\\
  \hline
0.2178& 40.00&52.95&19.36&7.31 &10.31&7.94&
0.2933& 40.00&42.08&15.46&7.26 &11.17&9.63  \\
 \hline
 0.2135&38.00 &46.83&19.85&7.62 &9.74&7.64&
0.2632&30.00 &31.51&20.15&7.70 &8.14&6.90 \\
 \hline
 0.1910& 40.00&46.47&22.22&7.70 &9.94&7.78&
0.2105& 31.00&32.31&25.02&7.67 &7.92&6.56 \\
  \hline
0.1881&44.00 &52.93&21.35&7.62 &10.89&8.42&
0.2000&36.00 &41.89&20.65&7.67 &9.06&7.26 \\
 \hline
 0.1683&44.00 &52.53&22.72&7.68 &10.56&8.03&
0.1600&37.00 &41.52&22.42&7.63 &8.75&6.86 \\
  \hline
 \end{tabular}
 \caption{Radii for ISCO and mass shedding for the end of the sequences
 with mass ratios $q$ shown.  $RL_E$ is  the Roche Lobe radius according
 to Eggleton  \cite{eggleton} and  $RL_{PN}$ is the Roche Lobe radius
 calculated n the second Post Newtonian approximation according to
 Ratkovic et. al.  \cite{ratkovic}.  The table shows also the proper radius of the NS $R_{NS}$ and the smallest separation of each sequence with mass ratio $q$ which converges. }
 \label{table:summary}
 \end{table*}

Our analysis is only indicative, we know that the Roche Lobe and the NS
are not spherical since at small  separation the  surface of the Neutron
star is flattened either by effects of its proximity to the black hole
and by  its rotation. In addition, our description is not fully
relativistic,  however, for the set of parameters used in our
simulations,  the present study indicates that the dynamics of the
mergers could follow a track which goes beyond the
\textit{plunge} or \textit{tidal disruption} ones. A full time evolution of such a configuration would be necessary to do in order to confirm such hypothesis, but this will be the content of a following paper.

\begin{acknowledgments}

	A. Lanza  gratefully acknowledges SISSA continuos support and A.Bressan for
	stimulating  discussions.  The computations were performed by using
	SISSA Ulysses  cluster and CINECA Galileo100 under the HPC
	Collaboration Agreement between SISSA and CINECA.
	\end{acknowledgments}

\appendix
\section{Results for the GW200105 event}\label{GW200105} In this Appendix
we present the results for the simulation of the GW200105 event.  The
figures and tables are similar to those presented in the main text but
they refer to a different  QE sequences aimed to simulate the GW200115
event.

\begin{figure*}[!ht]
	\begin{minipage}[b]{0.45\linewidth}
		\centering
\includegraphics[scale=0.45]{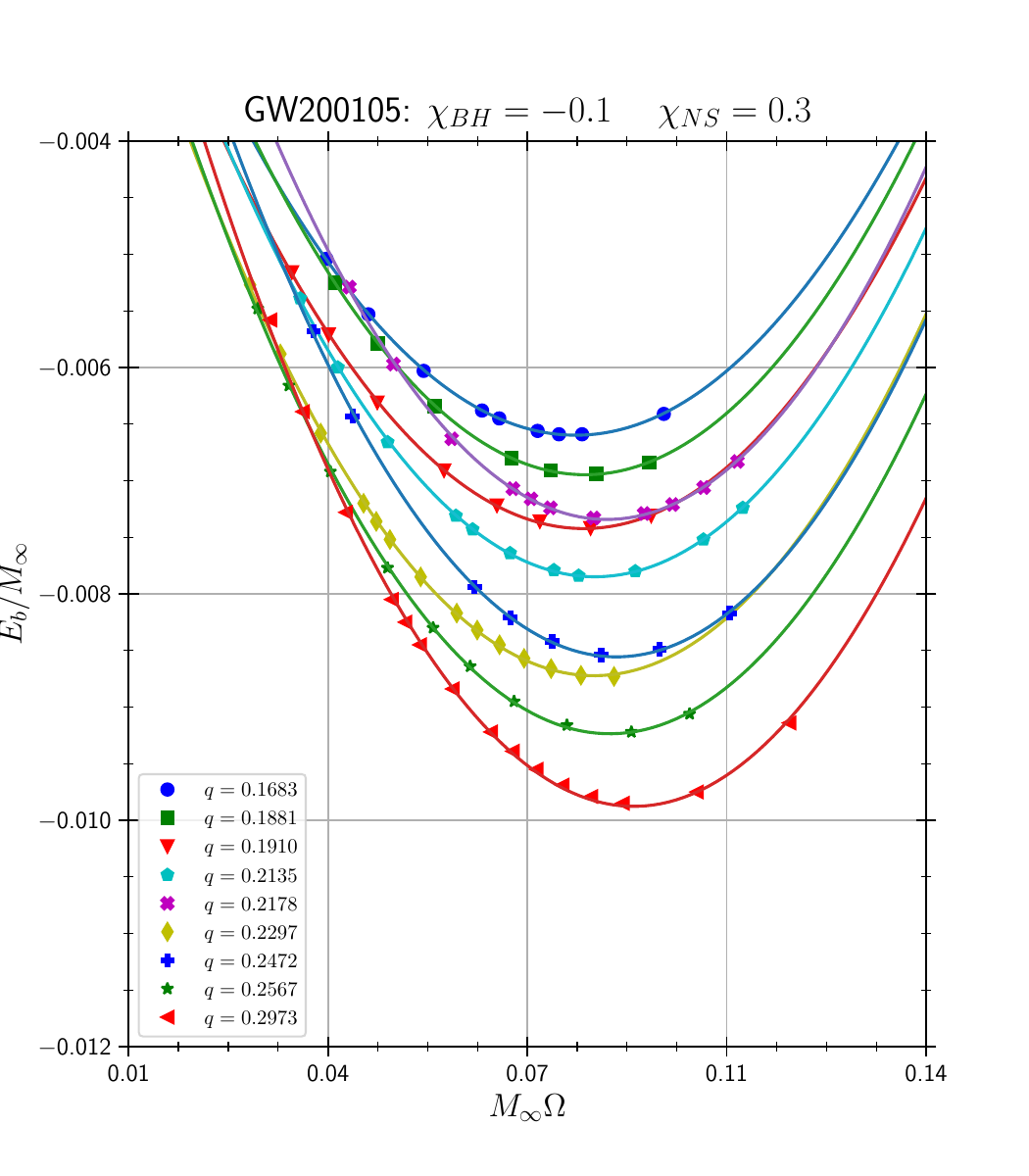}
 	\end{minipage}\hfill
 	\begin{minipage}[b]{0.50\linewidth}
		\centering
\includegraphics[scale=0.45]{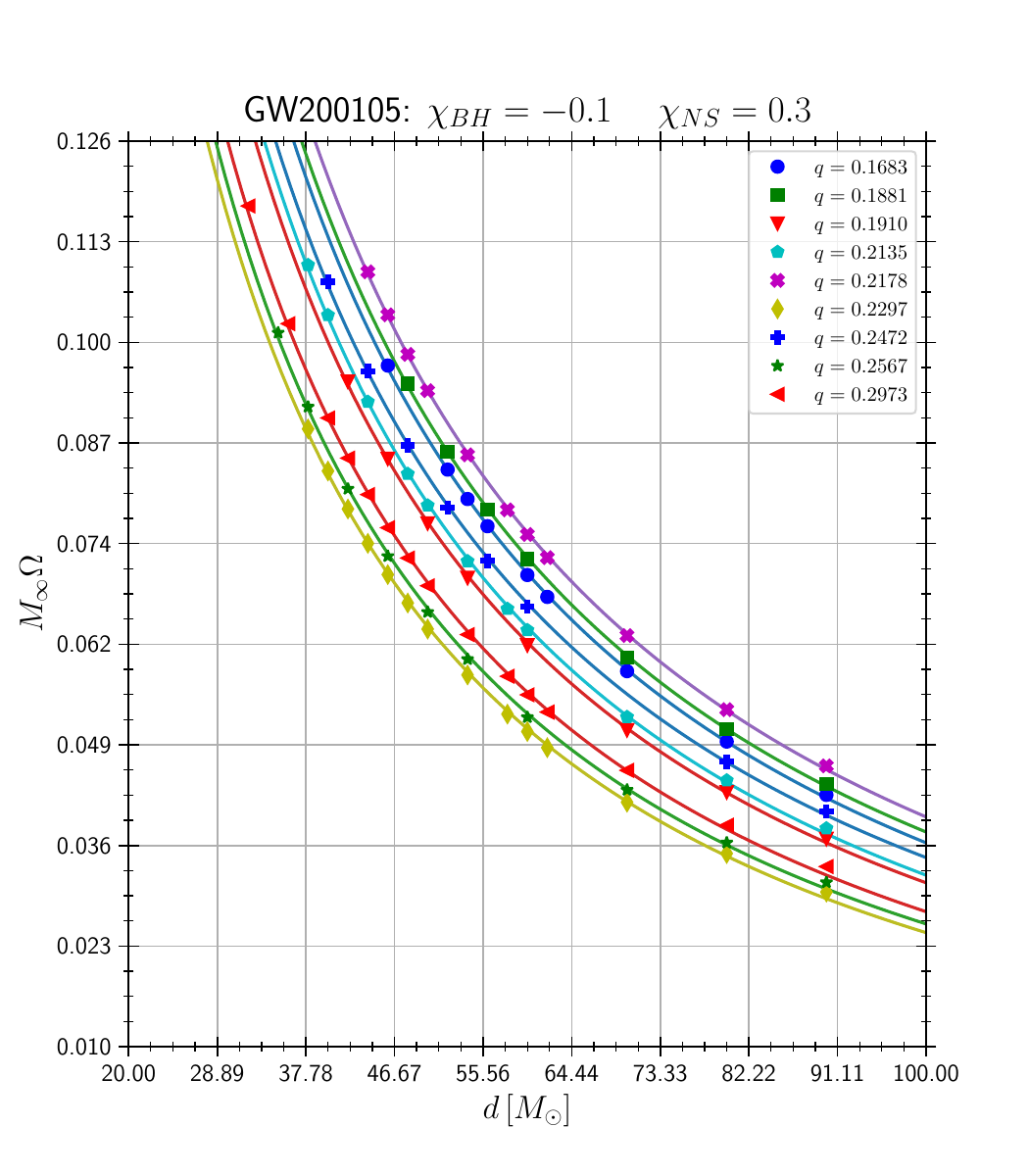}

	\end{minipage}
	\caption{\textit{Left panel:} Binding energy versus orbital
	angular velocity for different mass ratio and spin $\chi_{NS}=0.3$ for
	the neutron star and $\chi_{BH}=-0.1$ for the black hole for sequences
	simulating GW200105 event.  The values of the single masses determining
	the values of the parameter $q$  are given in Table
	\ref{table:GW2001X5}.  \textit{Right panel: }Orbital angular velocity
	versus binary separation. In both panels, different colored symbols are
	the values coming from the calculated sequences, solid lines of the
	same color report the fit with the functional form given by Eq.
	(\ref{eq:hyperbola}).}
	\label{fig:Eb_Omega_fit_gw105}
 \end{figure*}

Fig.  \ref{fig:Eb_Omega_fit_gw105} shows the fitting of the binding
energy versus the orbital angular velocity (\textit{left panel}) with a
parabola and the fitting of the orbital angular velocity versus the
separation  (\textit{right panel}).

In Table \ref{table: fit_params_GW105} we show the fitting parameters
 $b_i$ for $i=1,2,3$ and the values of $M_\infty\Omega_{ISCO}$.Also, the
 fitting parameters $a$ and $b$  and the values of $d_{ISCO}/M_\infty$
 and $d_{ISCO}$ are shown for each sequence simulating the event
 GW200105.

\begin{table*}[!ht]
    \centering
    \begin{tabular}{|c|c|c|c|c|c|c|c|}
    \hline
     \multicolumn{8}{|c|}{ \textbf{GW200105 }}  \\
  \hline
   \multicolumn{8}{|c|}{$\chi_{BH}=-0.1$ \quad $\chi_{NS}=0.3$} \\
  \hline
   q & b1 & b2 & b3 & $(M_{\infty}\Omega)_{ISCO}$ & a & b &$ d_{ISCO}$ \\ \hline
        0.2973 & 0.000348 & -0.2214 & 1.1989 & 0.0924 & 4.2161 & -0.0149 & 39.32 \\
        0.2567 & -0.000404 & -0.1997 & 1.1288 & 0.0885 & 4.0423 & -0.0147 & 39.19 \\
        0.2472 & 0.000806 & -0.2092 & 1.1686 & 0.0895 & 4.8854 & -0.0146 & 46.91 \\
        0.2297 & -0.000687 & -0.1872 & 1.0898 & 0.0859 & 3.9195 & -0.0146 & 39.02 \\
        0.2178 & 0.001556 & -0.2023 & 1.1497 & 0.0880 & 5.4637 & -0.0152 & 52.95 \\
        0.2135 & -0.000029 & -0.1818 & 1.0567 & 0.0860 & 4.7620 & -0.0157 & 46.83 \\
        0.1910 & -0.000377 & -0.1674 & 0.9940 & 0.0842 & 4.6186 & -0.0152 & 46.47 \\
        0.1881 & 0.000315 & -0.1721 & 1.0193 & 0.0844 & 5.2745 & -0.0153 & 52.93 \\
        0.1683 & -0.000147 & -0.1556 & 0.9388 & 0.0829 & 5.1769 & -0.0157 & 52.53 \\ \hline

    \end{tabular}
   \caption{Fitting parameters for the binding energy $Eb/M_{\infty}$
   ($b_1, B_2, b_3$) and the value of the orbital angular velocity at its
   minimum $M_\infty\Omega_{ISCO}$.  Also, the table show the fitting
   parameters for the orbital angular velocity $M_\infty\Omega$ and the value of
   the separation $d_{ISCO}[M_\odot]$ corresponding to $M_\infty\Omega_{ISCO}$. }
\label{table: fit_params_GW105}
\end{table*}

In Table \ref{table:tidal_105} we show  the same quantities  showed in
Table \ref{table:tidal_115} for the sequences relative to GW200115. For
all sequences the mass shedding radius is found at a location  less  than
the ISCO radius.

\begin{table*}[!ht]
\centering
\begin{tabular}{|c|c|c|c|c|c|}
\hline
  \multicolumn{6}{|c|}{ \textbf{GW200105 }}  \\
  \hline
   \multicolumn{6}{|c|}{$\chi_{BH}=-0.1$ \quad $\chi_{NS}=0.3$} \\
  \hline
$q$&$d[M_\odot]$&$R_{NS}$&$r_{ISCO}$&$r_{MS}$&$1/\mathcal{\kappa}$     \\
\hline
0.2973&32.00&7.34&39.32&17.62&1.27    \\
 \hline
0.2567&38.00&7.64&39.19&20.15&1.33   \\
 \hline
0.2472&38.00&7.33&46.91&18.71&1.27   \\
 \hline
0.2297&36.00&7.72&39.02&21.38&1.35 \\
  \hline
0.2178&40.00&7.31&52.95&19.36& 1.26  \\
 \hline
0.2135&38.00&7.62&46.83&19.85& 1.31   \\
 \hline
0.1910&40.00&7.70&46.47&22.22& 1.32   \\
  \hline
0.1881&44.00&7.62&52.93&21.35& 1.27 \\
 \hline
0.1683&44.00&7.68&52.53&22.72&1.30   \\
  \hline
 \end{tabular}
 \caption{$d[M_\odot]$ is the minimum separation along the sequence with a given
 mass ratio which converged. $R_{NS}$ is the Areal radius of the Neutron
 star, that is the proper radius of the NS as measured on the stellar
 surface.  $r_{MS}$ is the mass shedding radius as given by Eq.
 (\ref{eq:rMS}) and $1/\kappa$, given by Eq. (\ref{eq:kappa}). All the radius
 are in geometrical units.}
 \label{table:tidal_105}
 \end{table*}

\begin{figure*}[!ht]
\centering
{\includegraphics[width=.95\linewidth]{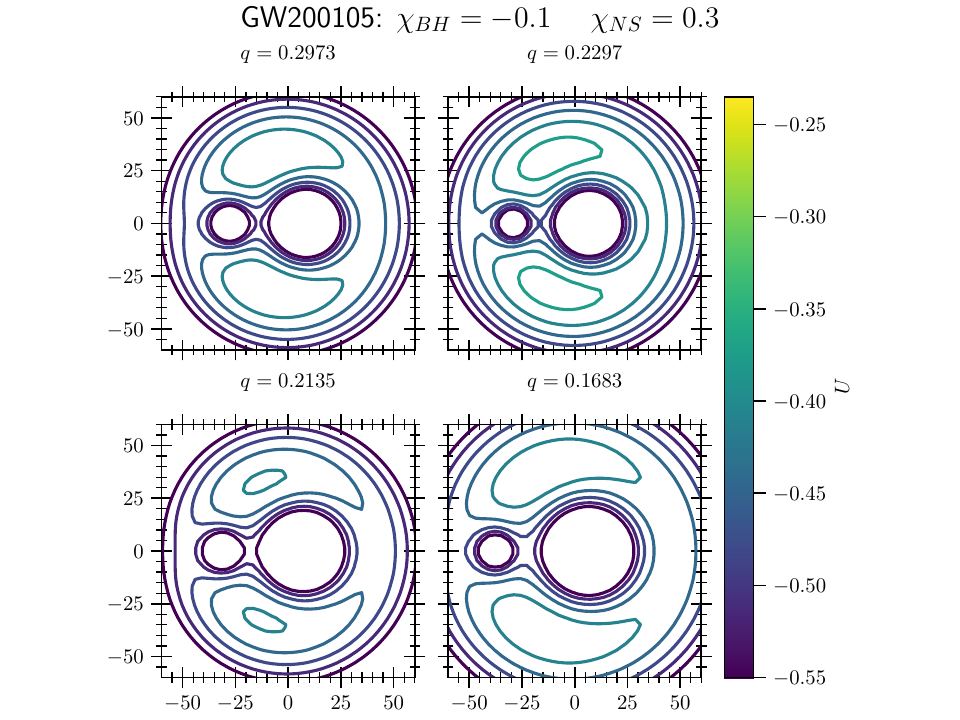}}
\caption{Reduced Newtonian Effective Potential surfaces for four  QE
sequences with different mass ratios  in a frame with origin at the
center of mass. The plots refer to the last converged solution of each
sequence to simulate GW200105 with spin $\chi_{NS}=0.3$ and
$\chi_{BH}=-0.1$.}
 \label{fig:equi105}
\end{figure*}

Fig. \ref{fig:equi105} shows the reduced Newtonian Effective Potential for four different mass ratios. The plots refer to the last converged solution of each sequence to simulation of the GW200105 event with spin $\chi_{NS}=0.3$ and
$\chi_{BH}=-0.1$.

\begin{figure*}[!ht]
\centering
{\includegraphics[scale=0.95]{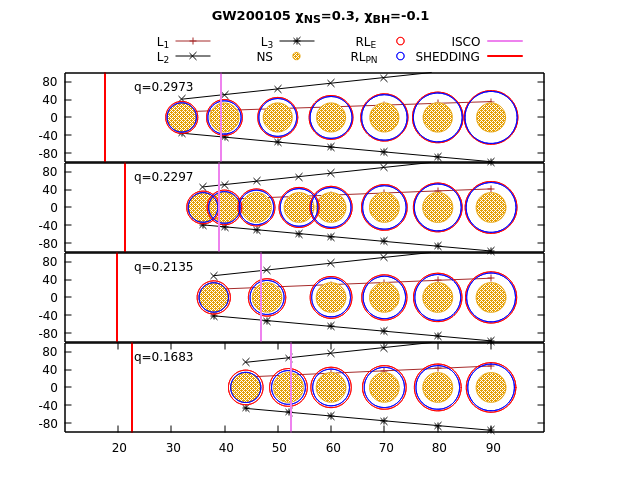}} 
\caption{Radii of reduced
Newtonian Effective Potential surfaces for different separation compared
with the relative positions of the collinear Lagrange points $L_1$, $L_2$
and $L_3$.  The filled circle represents the Neutron star with the proper
radius given by the QE sequence,  the red circunference is a circle whose
radius is given by the approximated Eggleton's radius of the Roche lobe
\cite{eggleton}, the radius of the blue circle has been calculated by
using the Post-Newtonian formula given in \cite{ratkovic}. Violet
vertical line corresponds to the ISCO radius and the red one to the
radius at which mass shedding starts to be important. These cases
correspond to sequences that simulate GW200105 event with spin
$\chi_{NS}=0.3$ and $\chi_{BH}=-0.1$ and different mass ratios.}
\label{fig:radii105}
\end{figure*}
In Fig. \ref{fig:radii105} we show the comparison between the radii of the reduced Newtonian Effective Potential surfaces for different separation compared
with the relative positions of the collinear Lagrange points $L_1$, $L_2$
and $L_3$.  The filled circle represents the Neutron star with the proper
radius given by the QE sequence,  the red circunference is a circle whose
radius is given by the approximated Eggleton's radius of the Roche lobe
\cite{eggleton}, the radius of the blue circle has been calculated by
using the Post-Newtonian formula given in \cite{ratkovic}. Violet
vertical line corresponds to the ISCO radius and the red one to the
radius at which mass shedding starts to be important. These cases
correspond to sequences that simulate GW200105 event with spin
$\chi_{NS}=0.3$ and $\chi_{BH}=-0.1$ and different mass ratios.
\clearpage

\bibliography{bibliography.bib}{}

\nocite{*}

\end{document}